\documentclass[lettersize,journal]{IEEEtran}
\usepackage{amsmath,amsfonts}
\usepackage{algorithmic}
\usepackage{algorithm}
\usepackage{array}
\usepackage[caption=false,font=scriptsize,labelfont=sf,textfont=sf]{subfig}
\usepackage{textcomp}
\usepackage{stfloats}
\usepackage{url}
\usepackage{verbatim}
\usepackage{graphicx}
\usepackage{cite}
\usepackage{booktabs}
\usepackage{makecell}
\usepackage{tabularx}

\usepackage[breaklinks,hidelinks,bookmarksdepth=3,bookmarksopen=true,bookmarksopenlevel=2,bookmarksnumbered=true]{hyperref}
\usepackage[capitalize]{cleveref}
\crefname{section}{Sec.}{Secs.}
\Crefname{section}{Section}{Sections}
\Crefname{table}{Table}{Tables}
\crefname{table}{Tab.}{Tabs.}
\allowdisplaybreaks[4]

\begin{document}

\title{A Novel Deep Learning-Based Coarse-to-Fine Frame Synchronization Method for OTFS Systems}

\author{Meiwen Men, Tao Zhou,~\IEEEmembership{Senior Member,~IEEE,} Kaifeng Bao, Zhiyang Guo, Yongning Qi,\\Liu Liu,~\IEEEmembership{Member,~IEEE,} and Bo Ai,~\IEEEmembership{Fellow,~IEEE}
\thanks{This work was supported by the National Natural Science Foundation of China under Grants 62571024, 62221001 and 62341127. (Corresponding author: Tao Zhou).}
\thanks{M. Men, T. Zhou, K. Bao, Y. Qi and L. Liu are with the School of Electronic and Information Engineering, Beijing Jiaotong University, Beijing 100044, China (e-mail: menmeiwen@bjtu.edu.cn; taozhou@bjtu.edu.cn; 23111017@bjtu.edu.cn; yongningqi@bjtu.edu.cn; liuliu@bjtu.edu.cn).}
\thanks{Z. Guo is with the School of Information Science and Technology, University of Science and Technology of China , Hefei, Anhui 230026, China (e-mail: guozhiyang@mail.ustc.edu.cn).}
\thanks{B. Ai is with the State Key Laboratory of Advanced Rail Autonomous Operation, Beijing 100044, China, and with the School of Electronics and Information Engineering, Beijing Jiaotong University, Beijing 100044, China (e-mail: boai@bjtu.edu.cn).}}

\maketitle

\begin{abstract}
Orthogonal time frequency space (OTFS) modulation is a robust candidate waveform for future wireless systems, particularly in high-mobility scenarios, as it effectively mitigates the impact of rapidly time-varying channels by mapping symbols in the delay-Doppler (DD) domain. However, accurate frame synchronization in OTFS systems remains a challenge due to the performance limitations of conventional algorithms. To address this, we propose a low-complexity synchronization method based on a coarse-to-fine deep residual network (ResNet) architecture. Unlike traditional approaches relying on high-overhead preamble structures, our method exploits the intrinsic periodic features of OTFS pilots in the delay-time (DT) domain to formulate synchronization as a hierarchical classification problem. Specifically, the proposed architecture employs a two-stage strategy to first narrow the search space and then pinpoint the precise symbol timing offset (STO), thereby significantly reducing computational complexity while maintaining high estimation accuracy. We construct a comprehensive simulation dataset incorporating diverse channel models and randomized STO to validate the method. Extensive simulation results demonstrate that the proposed method achieves robust signal start detection and superior accuracy compared to conventional benchmarks, particularly in low signal-to-noise ratio (SNR) regimes and high-mobility scenarios.
\end{abstract}


\section{Introduction}
\IEEEPARstart{O}{rthogonal} time frequency space (OTFS) modulation is a promising candidate waveform for future wireless communication systems due to its robustness to time-varying wireless channels and backward compatibility with orthogonal frequency division multiplexing (OFDM) \cite{ref32}. Traditional OFDM systems \cite{ref1} face severe degradation in high-mobility scenarios due to Doppler-induced inter-carrier interference (ICI). By performing symbol mapping in the delay-Doppler (DD) domain, OTFS represents the time-varying channel characteristics as sparse features. This not only reduces channel estimation complexity but also effectively mitigates the Doppler effect encountered in high-mobility scenarios

Frame synchronization, as a crucial step in wireless transmission for recovering packet data, is of great significance to both the receiver system and the entire communication system \cite{ref33}. Eliminating symbol timing offset (STO) to achieve precise frame synchronization is a fundamental requirement in contemporary wireless communication networks. While traditional synchronization algorithms, especially those targeting OFDM systems, can be  applied to OTFS in static environments, their performance significantly degrades in high-mobility scenarios, necessitating new strategies to enhance system performance \cite{ref27}. Therefore, exploring novel synchronization schemes to overcome the limitations of traditional methods has become a core issue in advancing the practical application of OTFS technology. 

Conventional frame synchronization methods have been extensively studied, especially within the context of OFDM systems. These methods typically rely on preambles or the inherent structural properties of the signal. For instance, the classic algorithm proposed in \cite{ref2} utilizes a special symbol composed of two identical halves, enabling timing estimation by identifying the peak of an autocorrelation function. Although subsequent research \cite{ref3} and \cite{ref4} improved timing precision by refining the preamble structure, these conventional methods remain vulnerable in high-mobility scenarios, despite their effectiveness in static or low-mobility environments. The resulting fast time-varying channels cause severe Doppler spread, which disrupts the orthogonality between subcarriers and leads to a sharp decline in the performance of correlation-based algorithms, rendering them inadequate for the stringent demands of future dynamic scenarios.

\IEEEpubidadjcol

To address these challenges, researchers have begun to develop synchronization schemes specifically tailored for OTFS systems. Some approaches incorporate preambles, such as the random-access preamble designed for the OTFS uplink in \cite{ref5}, though its timing feedback may become outdated. Similarly, a preamble-based method for the downlink was proposed in \cite{ref6}; however, its performance degrades with large frame sizes, while small frames incur high overhead \cite{ref7}. To reduce this overhead, recent studies have shifted towards exploiting the pilots embedded within the OTFS frame for channel estimation. For example, the methods in \cite{ref8} and \cite{ref9} propose a preamble-less algorithm that determines synchronization position by calculating the sparsity of the received signal in the DD domain, but its performance is suboptimal at low signal-to-noise ratios (SNRs). Another prominent category of synchronization methods in \cite{ref10} and \cite{ref11} leverages the periodic characteristics of embedded pilots in the delay-time (DT) domain. Although these techniques do not require additional training overhead and have no limitations in the estimation range, they often require the computation of two-dimensional (2D) autocorrelation functions. This leads to prohibitive computational complexity that impedes real-time implementation. Furthermore, as noted in \cite{ref12}, interference from adjacent frames can compromise the synchronization accuracy of the current frame. Consequently, conventional OTFS synchronization methods struggle to achieve an ideal balance among overhead, robustness, and computational complexity.

In recent years, deep learning (DL) has emerged as a powerful new paradigm for redesigning the physical layer, as introduced in foundational works like \cite{ref13}. Researchers have explored end-to-end system learning through autoencoders \cite{ref14} and even learning over unknown physical channels using adversarial networks \cite{ref15}. This trend has naturally extended to solving specific, challenging tasks such as synchronization. Researchers began to model frame synchronization as a classification or regression problem. For instance, \cite{ref16} demonstrated a practical over-the-air system where a neural network was explicitly used for frame synchronization in a continuous data stream. Other works have applied various architectures in OFDM systems, such as convolutional neural networks (CNNs) for classification \cite{ref16}, regression \cite{ref17}, or as fully convolutional networks (FCNs) as nonlinear deep filters \cite{ref18}, and extreme learning machines (ELMs) for fine timing and frequency offset estimation \cite{ref19}. However, these existing DL-based methods have notable shortcomings. Most models are validated under relatively simple channel conditions, and some schemes, like \cite{ref18}, still depend on extra preambles, failing to completely eliminate the overhead of traditional methods. Low-overhead, low-complexity DL-based solutions for OTFS frame synchronization in high-mobility channels remain underexplored.

As discussed above, there is currently no perfect solution for DL-based methods in synchronization tasks, especially for OTFS systems. To bridge these research gaps, this paper investigates the DL-based frame synchronization method for OTFS systems. The major contributions and novelties of this paper are as follows.

\begin{enumerate}
\item{We establish a comprehensive OTFS system model to investigate the detrimental effects of STO and formulate the frame synchronization task as a classification problem. This formulation reveals the prohibitive complexity of a one-stage classifier, motivating the novel hierarchical approach presented in this work.}
\item{We propose a low-complexity coarse-to-fine frame synchronization method based on a residual network (ResNet) architecture. This non-data-aided approach eliminates reliance on traditional preambles by leveraging the periodic features of embedded pilots. It employs a two-stage classification strategy to first narrow the search space and then pinpoint the exact STO, significantly reducing computational complexity while maintaining high precision.}
\item{We conduct a comprehensive performance evaluation by constructing a simulation dataset encompassing diverse and challenging channel conditions. Extensive simulations validate our method's superior accuracy, low estimation error, and robustness against conventional algorithms and other DL architectures, particularly in low SNR and high-mobility scenarios. A complexity analysis further confirms the method's efficiency.}
\end{enumerate}

The remainder of this paper is outlined as follows. Section II establishes the OTFS system model and formulates the frame synchronization task as a classification problem. In Section III, the novel coarse-to-fine frame synchronization method based on ResNet is proposed, detailing the network architecture and implementation strategy. Then, the dataset generation process, performance evaluation results, and complexity analysis are presented in Section IV. Finally, conclusions are drawn in Section V.

\section{Problem Formulation for Frame Synchronization in OTFS Systems}
\subsection{System Model}
\begin{figure*}[!t]
\centering
\includegraphics[width=0.8\linewidth]{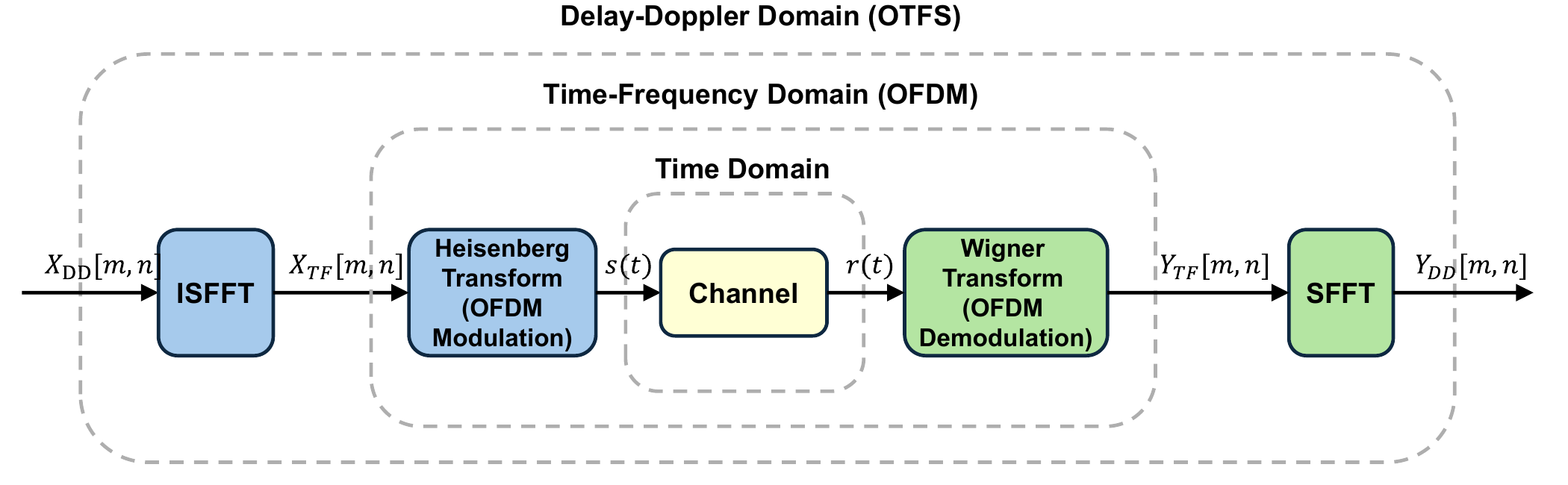}
\caption{OTFS system block diagram.}
\label{fig_1}
\end{figure*}

As shown in \cref{fig_1}, OTFS modulation involves a series of two-dimensional (2D) transformations at both the transmitter and receiver \cite{ref20}. The transmitter first maps the modulated symbols $\mathbf{X}_{\text{DD}}[m,n]$ (for $m=0,\dots,M-1$ and $n=0,\dots,N-1$) on the DD grid to the time-frequency (TF) domain through inverse symplectic finite Fourier transform (ISFFT) and windowing operations, yielding $\mathbf{X}_{\text{TF}}[m,n]$. Subsequently, $\mathbf{X}_{\text{TF}}[m,n]$ is converted into a continuous time-domain signal $s(t)$ via Heisenberg transform. At the receiver, the demodulation process initiates by reshaping the received signal $r(t)$ into an $M \times N$ DT domain $r[m,n]$. Note that the Wigner transform serves as the inverse of the Heisenberg transform, mapping the signal back to the TF domain. 

The complex baseband channel impulse response $h(\tau, \nu)$ describes the channel's response as a pulse with delay $\tau $ and Doppler shift $\nu$. In this DD domain representation, $h(\tau, \nu)$ plays a role analogous to the traditional impulse response in time domain $h(t)$, but with a distinct advantage: its compact structure. Specifically, only the Doppler shift $\nu$ and delay $\tau$ are required to accurately describe the path information. Conversely, in the time domain, due to the time-varying and frequency-selective nature caused by mobility and multipath, the pilot spreads throughout the entire time domain, which requires a larger volume of data to describe the channel state information, and the prominence of the pilot is also weakened due to the spread \cite{ref21}.

\begin{figure}[!t]
\centering
\includegraphics[width=1.0\linewidth]{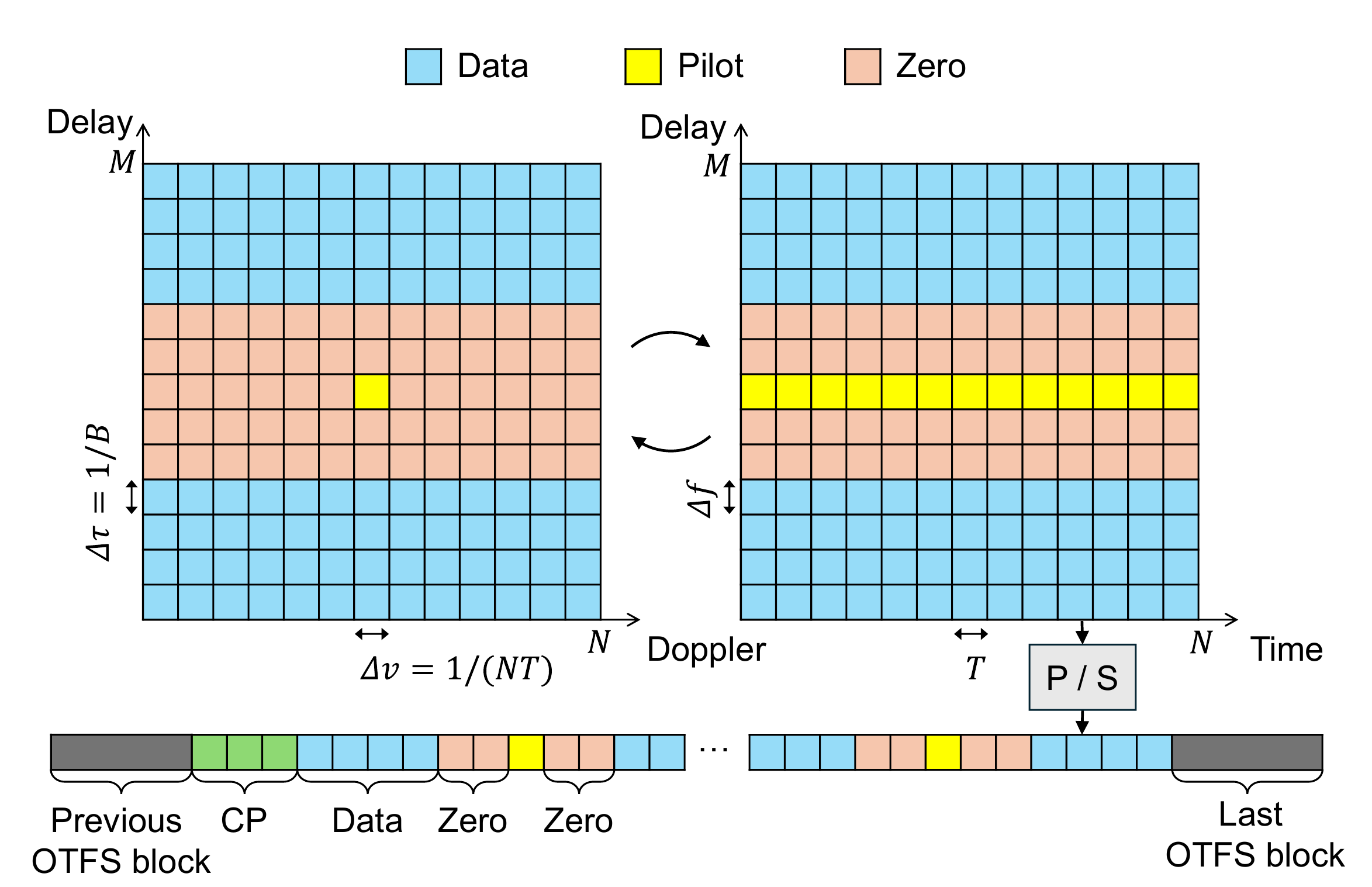}
\caption{DD domain signal with embedded pilot and its counterpart in the DT domain.}
\label{fig_2}
\end{figure}

\cref{fig_2} shows the embedded pilot signal scheme used in this paper. Each OTFS block contains $M \times N$ (for some integers $M,N>0$) sample points. A pulse pilot is inserted at $(m_p, n_p)$ in the DD grid, surrounded by guard symbols, which are usually set to zero. To further reduce overhead, only one cyclic prefix is used at the beginning of each OTFS block, rather than multiple cyclic prefixes in each block. This scheme is the same as the channel estimation pattern used in \cite{ref8}, requiring guard symbols in the delay dimension to avoid interference between pilot and data symbols at the receiver. Then, the DD domain signal is transformed to the time domain for propagation through OTFS modulation.

\begin{figure}[!t]
\centering
\subfloat[]{\includegraphics[width=0.5\linewidth]{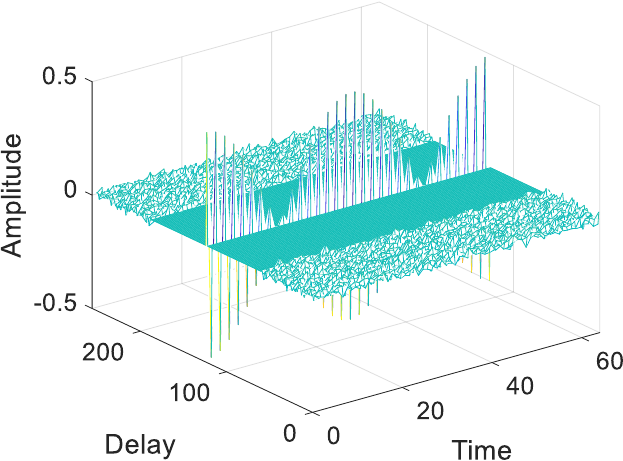}\label{fig3_first_case}}%
\hfil
\subfloat[]{\includegraphics[width=0.5\linewidth]{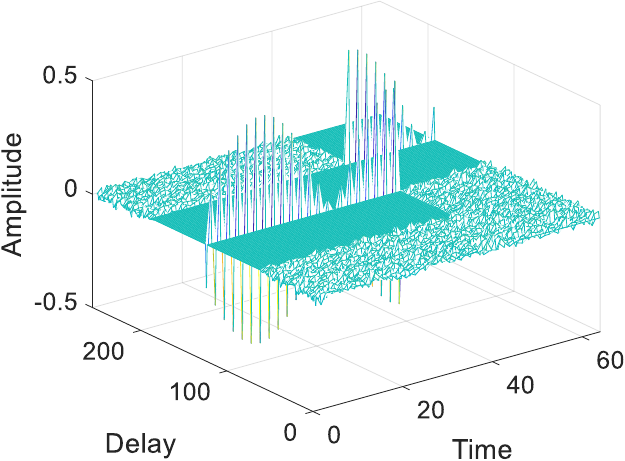}\label{fig3_second_case}}%
\caption{The real part of the signal in DT grid. (a) Transmitted signal. (b) Received signal.}
\label{fig_3}
\end{figure}

As depicted in \cref{fig_3}, the pilots exhibit a periodic structure within the DT domain. \cref{fig_3}~\subref{fig3_first_case} shows the real part of the transmitted signal in the DT domain, clearly illustrating this periodic arrangement of the pilots. \cref{fig_3}~\subref{fig3_second_case} presents the real part of the received signal in the DT domain under a scenario only with STO. Here, it is evident that the rows corresponding to the pilots are shifted along both the delay and time dimensions.

\subsection{Problem Formulation}
\begin{figure}[!t]
\centering
\includegraphics[width=1.0\linewidth]{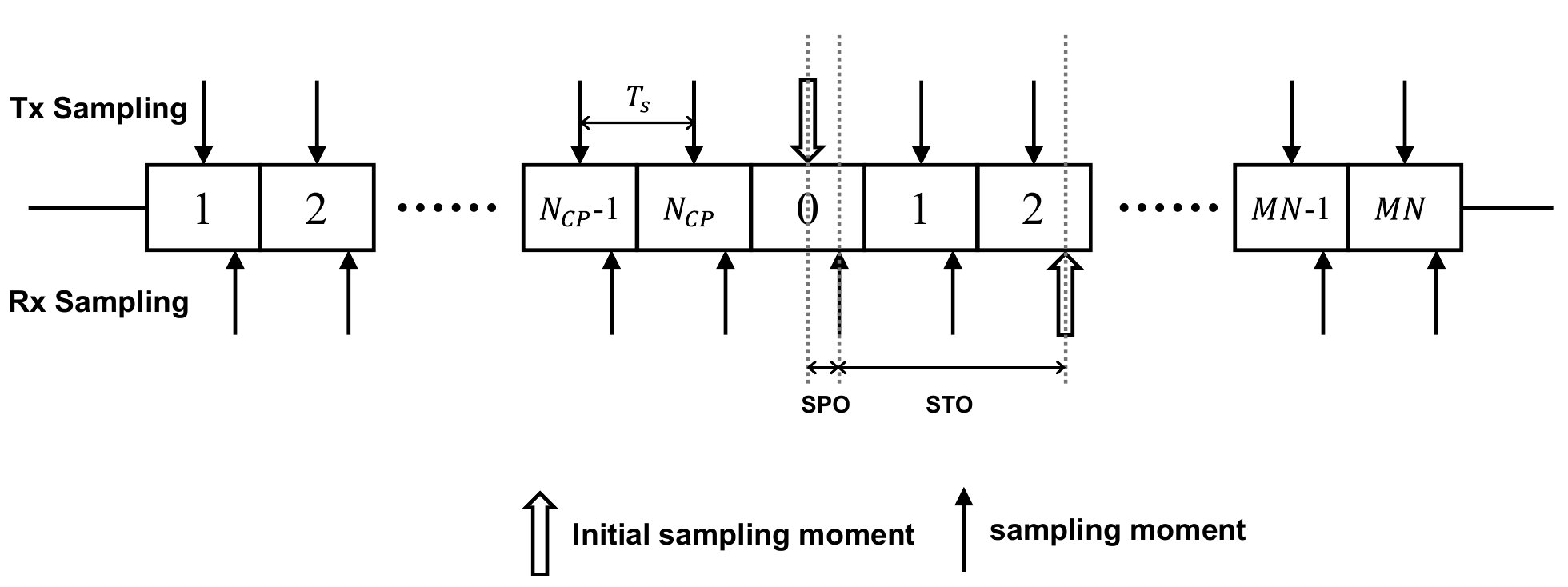}
\caption{Time domain signal sampling.}
\label{fig_4}
\end{figure}

As illustrated in \cref{fig_4}, timing offset is primarily composed of two components. One is the STO, denoted as $\operatorname{Int}\{\theta \}$, which results from the misalignment between the true and estimated starting positions of the discrete Fourier transform (DFT) window. The other component is the sampling phase offset (SPO), caused by asynchronous sampling clocks at the transmitter and receiver. The SPO represents the fractional timing offset $\operatorname{Frac}\{\theta \}$ \cite{ref23}.

To ensure accurate symbol recovery in the DD domain via an $M\times N$ points DFT, the receiver in an OTFS system must perform symbol timing synchronization by detecting the start of each symbol \cite{ref24}. Furthermore, when STO exists, it will cause the main path of the received signal to shift to other paths, which may lead to different degrees of ICI and inter-symbol interference (ISI) \cite{ref25,ref26}, and destroy the orthogonality of the OTFS signal. For an OTFS system, if the initial position of the first path cannot be found, there is a risk of channel estimation errors and signal detection failures \cite{ref28}.

Based on the structure of the OTFS block, the integer part of the timing offset is decomposed into $\operatorname{Int}\{\theta \}={{\theta }_{\text{d}}}+M{{\theta }_{\text{t}}}$, where ${{\theta }_{\text{d}}}$ and ${{\theta }_{t}}$ represent the offsets in the delay and time dimensions, respectively. The fundamental challenge of frame synchronization in OTFS systems is to accurately estimate  $\operatorname{Int}\{\theta \}$ from a received time-domain signal sequence $r[k]$. This sequence contains the desired OTFS frame corrupted by noise and channel effects; therefore, the goal is to find the precise starting sample of the frame. The received signal can be expressed as
\begin{equation}
\label{eq_1}
r[k] = \sum\limits_{i = 0}^{B - 1} {\sum\limits_{l = 0}^{{N_s} - 1} {h[l,k]s[k - l - \operatorname{Int}\{\theta \}  - i{N_s}]} }  + \eta [k],
\end{equation}
where $\eta[k]$ is the complex additive white Gaussian noise (AWGN) with the variance $\sigma^2$. Let the transmitted time-domain signal for a single OTFS frame block with cyclic prefix (CP) be represented by a sequence $s[k]$ of length $N_s = MN + L_{CP}$, $B$ is the number of OTFS blocks in each data frame, and $h[l,k]$ is the channel response at the delay-tap $l$ and sample $k$.

Due to the unknown timing offset, the receiver captures a signal window of the same length, which is a cyclically shifted version of the ideal signal. The task is to design an estimator such that the estimated $\hat{\theta}$ accurately approximates the true offset $\theta$ using only the received signal $r[k]$.

OTFS system synchronization schemes fall into two major categories: conventional techniques and DL-based techniques. Conventional algorithms refer to methods that perform induction or processing using linear mathematical models. Depending on the input data used, they can be additionally categorized into two types: data-aided schemes use training sequences or preambles known at the receiver for signal estimation, whereas non-data-aided schemes exploit redundant information contained within the received data sequence, including the CP and virtual carrier, to estimate the signal \cite{ref29}.

Data-aided synchronization typically relies on traditional correlation-based methods, which prepend preamble sequences with desirable correlation properties to the transmitted data. At the receiver, the starting position of the packet is identified by correlating the received signal with the known preamble or by exploiting the preamble's autocorrelation properties. This presents a key trade-off: while a longer preamble enhances detection performance, it does so at the cost of greater transmission overhead and power consumption \cite{ref22}.

Currently, there is a 2D autocorrelation algorithm based on embedded pilot for non-data-aided methods for OTFS signals \cite{ref10}. This approach involves computing the autocorrelation of the 2D signal grid row by row to obtain a 2D autocorrelation matrix $P[m,n]$ as
\begin{equation}
\label{eq_2}
P[m,n] = \sum\limits_{k = 0}^{N - 2} {{r^*}[m,n + k]r[m,n + k + 1]}.
\end{equation}
The maximum value index along the delay dimension (index where pilot is located) is found, and the index of the ${{\theta }_{d}}$ along the delay dimension is then used to deduce. Along the time dimension, since the pilot has a periodic structure after inverse discrete Fourier transform (IDFT) across the Doppler axis, the maximum value point of the pilot row along the time dimension is the starting point of the time dimension for that frame, which is ${{\theta }_t}$. Finally, the initial position of the signal can be obtained based on $\operatorname{Int}\{\theta \}={{\theta }_{\text{d}}}+M{{\theta }_{\text{t}}}$. In the following experiments, both methods will be used as baselines.

In contrast to these traditional estimation problems, we reformulate frame synchronization as a classification task. This paradigm is particularly well-suited for DL, as the input waveform and the desired single-peak output (corresponding to the correct timing offset) align perfectly with the architecture of classifiers such as CNN. A straightforward approach would be to design a single classifier that directly estimates the STO from all ${{N}_{s}}$ possible classes. However, for a typical OTFS system where ${{N}_{s}}$ can be very large, this one-stage classification approach leads to a model with an impractically large output layer. Such a model suffers from prohibitive computational complexity and a massive number of trainable parameters, rendering it difficult to train and unsuitable for real-time implementation.

Consequently, the frame synchronization problem is formulated as a classification task. Let $\mathbf{r} \in \mathbb{C}^{N_s}$ denote the vector representation of the received sequence $r[k]$. The goal is to estimate the STO $\hat{\theta}$ by maximizing the posterior probability derived by the DL network $f_{\Phi}$ with trainable parameters $\Phi$, as follows
\begin{equation}
\label{eq_problem_formulation}
\hat{\theta} = \operatorname*{argmax}_{i \in \{0, \dots, N_s-1\}} \, [f_{\Phi}(\mathbf{r})]_i,
\end{equation}
where $[f_{\Phi}(\mathbf{r})]_i$ denotes the probability that the $i$-th candidate corresponds to the true integer offset $\operatorname{Int}\{\theta\}$.

\section{A Novel DL-Based Coarse-to-Fine Frame Synchronization Method}
\subsection{Overview of the Proposed Method}
\begin{figure*}[!t]
\centering
\includegraphics[width=1.0\linewidth]{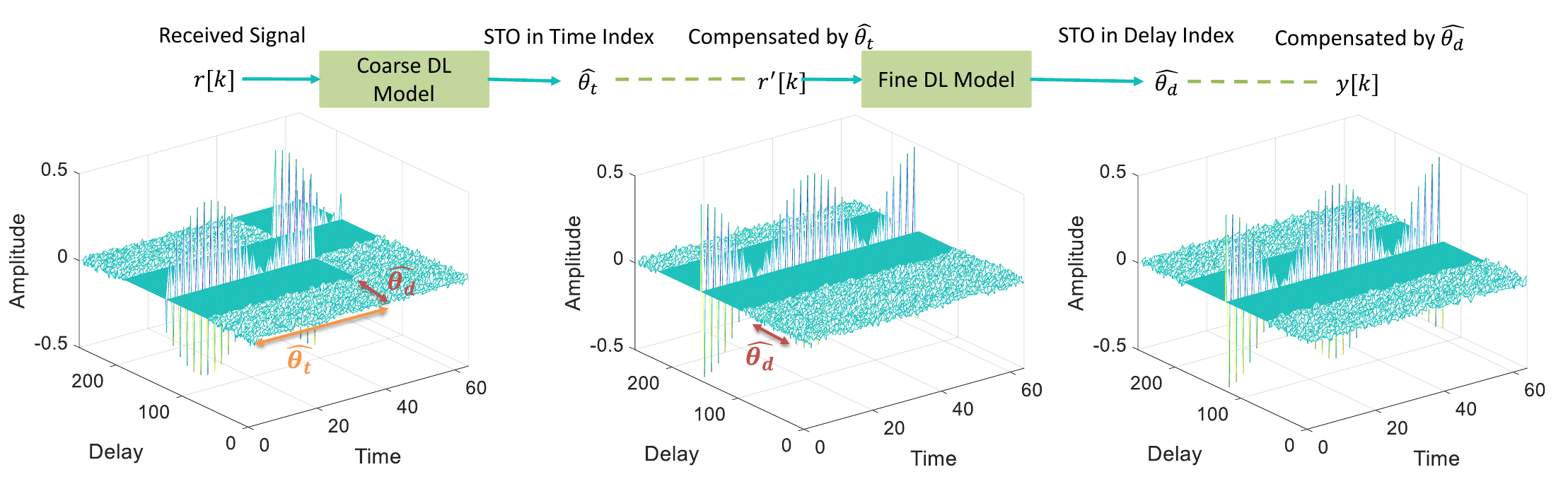}
\caption{Coarse-to-fine frame synchronization method.}
\label{fig_5}
\end{figure*}

To address the frame synchronization challenge in OTFS systems, we propose a novel method based on a two-stage, coarse-to-fine DL architecture. The overall workflow of this method is illustrated in \cref{fig_5}. The primary motivation for this approach is to mitigate the high computational complexity associated with a direct, one-stage classification. Directly identifying the start of a frame from all $MN$ possible sample positions would require a DL model with an excessively large output layer, leading to a substantial number of parameters and a demanding training process. Our coarse-to-fine strategy decomposes this large-scale problem into two smaller, more manageable classification tasks, which significantly reduces model complexity while maintaining high estimation accuracy.

The process begins with the coarse classification stage. The goal of this initial stage is to rapidly identify the approximate region of the signal that contains the frame's starting point. The received time-domain signal, a sequence of length $MN$, is fed into the first DL model. This model is trained to treat the entire sequence as being divided into $N$ contiguous segments, each of length $M$. Its task is to predict which of these $N$ segments contain the true start of the OTFS frame. The output of the coarse model is an integer index $\hat{\theta}_t$ from $0$ to $N-1$, which represents the estimated coarse location of the STO. This step effectively narrows down the search space for the precise offset from $MN$ possibilities to just $N$. 

Following the coarse estimation, the process moves to the fine classification stage. The output from the coarse model is first used to pre-process the signal. The original received signal is cyclically shifted so that the beginning of the segment identified in the coarse stage is aligned with the start of the processing window. This compensated signal $r'(t)=r(t-M{{\theta }_t})$ is then passed as input to a second, distinct DL model. The task of this fine model is to perform a high-resolution search within this pre-aligned, $M$-sample-long region. It is trained to classify among the $M$ possible sample positions to pinpoint the exact starting sample. The output of the fine model is an integer index $\hat{\theta}_d \in \{0, \dots, M-1\}$.

Finally, the complete STO is reconstructed by combining the outputs from both stages. The coarse segment index determines the large-scale shift, while the fine sample index provides the precise intra-segment position. This final, high-precision STO estimate is then used to apply a final cyclic shift to the original received signal, ensuring the receiver's sampling window is accurately aligned with the start of the OTFS frame for correct demodulation.

\subsection{ResNet Model for Coarse-to-Fine Frame Synchronization}

To implement this method, we employ a customized deep neural network architecture based on ResNet. This architecture is chosen for its ability to extract deep features, particularly for capturing the inherent periodic structures of OTFS pilots within the time-domain signal. By leveraging residual learning, the proposed architecture effectively mitigates the gradient vanishing problem, ensuring robust classification performance in both coarse and fine stages \cite{ref30}.

The fundamental building block of our network is the modified residual block (ResBlock), tailored for time-series signal processing \cite{ref31}. Unlike standard ResNets which utilize 2D convolutions for images, our model employs 1D convolutions to process the sequential nature of the received signal $r[k]$.
Let $x \in \mathbb{R}^{B \times C_{in} \times L}$ denote the input tensor to the $l$-th block, where $B$, $C_{in}$, and $L$ represent the batch size, number of input channels, and sequence length, respectively. The output $y$ of the ResBlock is defined as
\begin{equation}
\label{eq_3}
y = \sigma \left( \mathcal{F}(x, \{W_i\}) + \mathcal{H}(x) \right),
\end{equation}
where $\sigma(\cdot)$ denotes the rectified linear unit (ReLU) activation function and $\mathcal{F}(\cdot)$ represents the residual mapping to be learned, which consists of three stacked convolutional layers with kernel sizes of 7, 5, and 3, respectively. This combination of varying kernel sizes is designed to capture temporal features at different scales, effectively aggregating local and contextual information.

The term $\mathcal{H}(x)$ represents the shortcut connection, which performs an identity mapping to facilitate gradient backpropagation. To handle changes in channel dimensions between blocks, $\mathcal{H}(x)$ is formulated as
$$\mathcal{H}(x) = 
\begin{cases} 
x, & \text{if } C_{in} = C_{out} \\
W_s x, & \text{if } C_{in} \neq C_{out}
\end{cases}\quad,$$
where $W_s$ represents a $1 \times 1$ convolution operation used to align the input channel dimension $C_{in}$ with the output dimension $C_{out}$. Batch normalization (BN) is applied after each convolution and before the addition operation to accelerate convergence.

\begin{figure}[!t]
\centering
\includegraphics[width=1.0\linewidth]{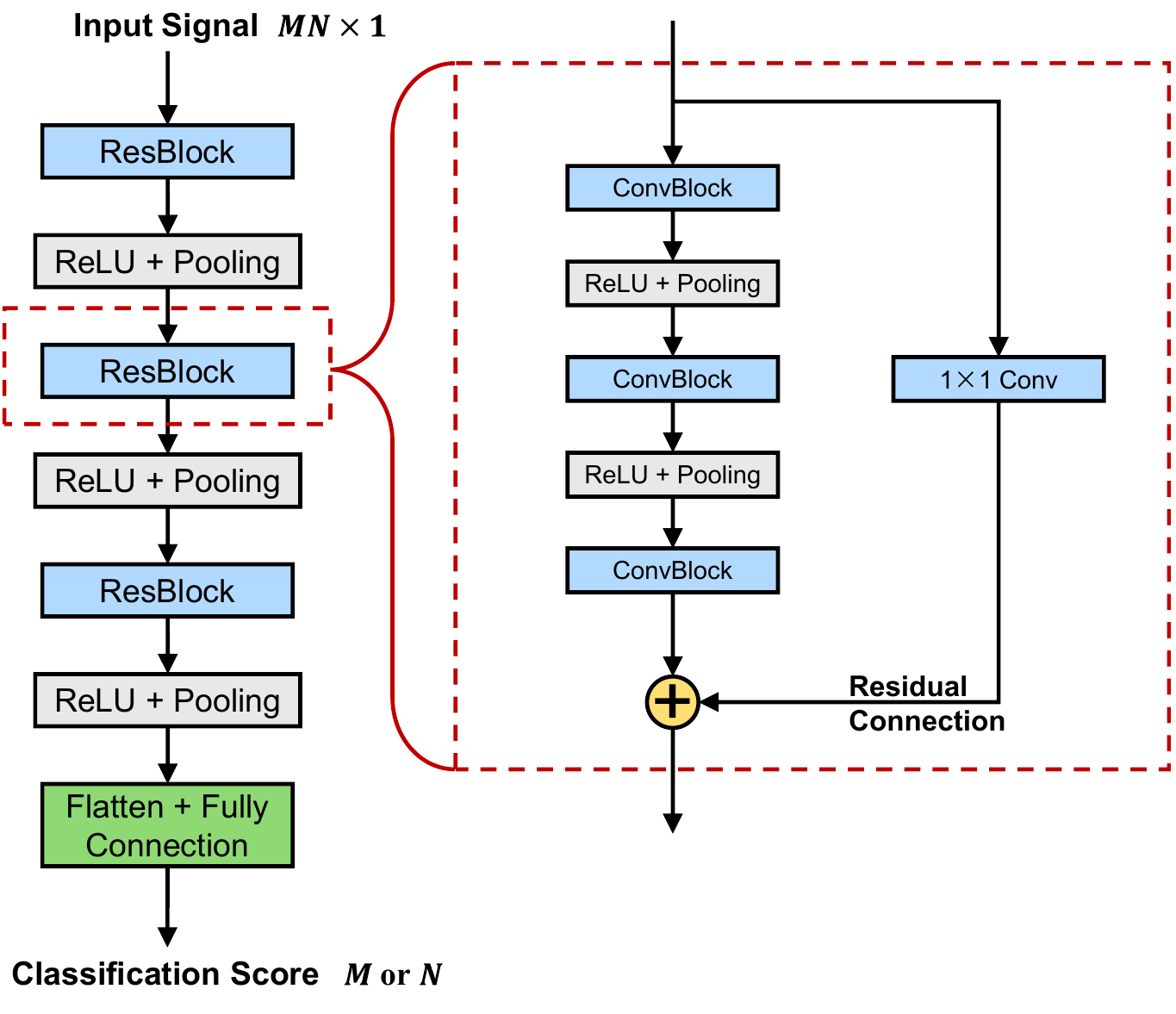}
\caption{Structure of the ResNet model for coarse-to-fine synchronization.}
\label{fig_6}
\end{figure}

As illustrated in \cref{fig_6}, the overall framework stacks multiple ResBlocks to form a deep feature extractor. The network progressively increases the number of channels (from 2 to 16) while reducing the temporal dimension via max pooling layers (kernel size 2, stride 2) after each stage. This hierarchical structure condenses the signal into a high-level feature vector, which is finally mapped to the synchronization output (segment index $\hat{\theta}_t$ or fine offset $\hat{\theta}_d$) via a fully connected layer.

In the coarse classification stage, the input of the model is a time series signal with length $MN$. The input signal first undergoes feature extraction through a network stacked with three ResBlocks and max pooling layers. After feature extraction, a fully connected layer maps the features to $N$ classes. Specifically, the input signal is divided into multiple segments of length $M$, and each segment corresponds to a label. The goal of this stage is to initially estimate the symbol timing offset of the signal and identify the segment where the possible frame header position is located. The input is a time series signal of length $MN$. Coarse classification is performed in $N$ dimensions, dividing the signal into $N$ segments, each containing $M$ sampling points. The output is the segment index where the estimated STO of the signal is located. Through coarse classification, the model can quickly locate the area where the frame header may exist, significantly reducing the search space for subsequent fine classification. The goal of this stage is to initially estimate the STO and narrow the search range to within one segment.

In the fine classification stage, the STO estimated in the coarse classification stage is used to compensate the original received signal and perform cyclic shifting to obtain a preliminarily aligned signal. The compensated signal's single-channel is also taken as the input for the fine classification model, with the length still being $MN$. The model performs more precise estimation on the segment identified in the coarse classification stage. At this point, the signal offset is limited to a smaller range $M$, and the model classifies the $M$ sampling points within each segment to determine the precise frame header position. The input is the signal after preliminary STO compensation. The $M$ sampling points within each segment are classified. The output is the precise sampling point index of the estimated frame header within the segment. Through this iterative approximation method, the model can achieve high-precision frame header position estimation while ensuring computational efficiency.

This coarse-to-fine structure utilizes the powerful feature extraction capability of ResNet and combines it with a progressive classification strategy to quickly locate the approximate range in the first stage and perform fine localization in the second stage, thereby achieving high-precision STO estimation while controlling computational complexity. 

\subsection{Model Implementation}
\subsubsection{Training Scheme}

In the coarse classification stage, the training input is the two-channel time-domain complex signal containing random STO (range covering the entire frame length), with a shape of [Batch Size, 2, $MN$], and the label is the segment index where the true STO of the signal is located.

The logits output from the model will be transformed into a probability distribution by softmax before calculating the cross-entropy loss, which is used as the loss function for training to quantify the difference between the model's estimated value and the true label by
\begin{equation}
\label{eq_4}
L =  - \sum\limits_{i = 1}^n {{y_i}} \log \left( {{e^{{x_i}}}/\sum\limits_{j = 1}^n {{e^{{x_j}}}} } \right),
\end{equation}
where ${{x}_{i}}$ is the value of the $i$ logits value of the class, which represents the network output of the $i$-class score, and ${{y}_{i}}$ is the true label.

We employ AdamW as the optimizer. By introducing the weight decay mechanism, AdamW can effectively alleviate overfitting problems, accelerate model convergence, and ensure a stable and efficient training process. We set the learning rate to 0.0001 to balance convergence speed and model accuracy and avoid the risk of overfitting.
The model training lasts for 500 epochs to ensure that the model can fully learn complex features in the data and reach a converged state. The batch size is set to 256. By processing a large number of samples in parallel, the training process is not only accelerated but also the generalization ability of the model is improved. During the training process, after each epoch, the model's performance is evaluated on the test set to monitor the learning progress in real-time and verify the model's performance under different conditions. After training is completed, the weights of the coarse classification model are saved.

In the fine classification stage, we first use the trained coarse model to predict the segment index for each sample. Based on the coarse classification estimation result, we perform cyclic shift compensation on the input signal. This compensated signal serves as the input for the fine classification model, where the label is the precise sampling point index within the corresponding segment.The hyperparameters (loss function, optimizer, learning rate, batch size, and epochs) remain consistent with the coarse stage. After training is completed, the weights of the fine classification model are saved.

Finally, in the evaluation stage, the received signal passes through the coarse classification model and the fine classification model sequentially. The final STO estimation value is obtained by combining the estimation results of the two stages, and the signal is compensated accordingly to complete frame synchronization.

\subsubsection{Hyper-parameters}

\begin{table}[!t]
\caption{Setting OF Model Parameters\label{table_1}}
\centering
\begin{tabular}{|l||c|}
\hline
Parameter & Value\\
\hline
Input length & $2 \times MN$\\
\hline
Number of channels & $(2, 4), (4, 16), (16, 16)$\\
\hline
Kernel size & $[7, 5, 3]$\\
\hline
Activation function & ReLU\\
\hline
Pooling layer kernel size and stride & MaxPool1d (2, 2)\\
\hline
Fully connected layer & $16 \times (MN / 8), M \text{ or } N$\\
\hline
\end{tabular}
\end{table}

\Cref{table_1} lists the key parameters of the used ResNet model. This convolutional kernel size has more advantages in feature extraction capabilities and can better capture signal details and adapt to time-series features of different scales. The channel number configuration is optimized through experiments, which not only ensures feature expression capability but also avoids excessive consumption of computing resources. These parameters are determined through theoretical analysis and a large number of experimental optimizations after comprehensively considering the characteristics of OTFS signals, the requirements of synchronization tasks, and the trade-off between model performance and complexity.

This parameter configuration is designed to fully utilize the advantages of the ResNet architecture, effectively capture the periodic pilot features in OTFS signals, and control model complexity through coarse-to-fine processing and pooling operations, ultimately achieving good overall performance in the STO estimation task.

\section{Performance Evaluation of the Proposed Method}
\subsection{Dataset Generation}
To evaluate the STO estimation and synchronization performance, we consider three channel conditions. For the multipath fading scenarios, the received signal is generated by first passing the transmitted time-domain signal through the respective fading channel conditions, and subsequently AWGN to match the target SNR. Specifically, the three channel models are defined as follows: Channel 1 (AWGN only), Channel 2 (Rayleigh channel with AWGN), and Channel 3 (extended vehicular A (EVA) channel with AWGN).

All simulations are conducted at a 3.35\,GHz center frequency and evaluated over a SNR range from $-20$\,dB to $26$\,dB. The Rayleigh channel is configured by a 3-path model with delays of $[0, 100, 200]$\,ns, corresponding average gains of $[0, -10, -15]$ dB, and a maximum Doppler shift of 1525\,Hz. The EVA channel utilizes a 9-path model configured according to 3GPP TS 36.104, with delays of $[0, 30, 150, 310, 370, 710, 1090, 1730, 2510]$\,ns, corresponding average gains of $[0, -1.5, -1.4, -3.6, -0.6, -9.1, -7.0,\\ -12, -16.9]$\,dB, and a maximum Doppler shift of 3051\,Hz.

The random value range of STO is $[-MN/2,MN/2)$. To simulate an arbitrary STO $\theta $, each signal frame is prepended and appended with randomly selected data segments. A window of length $MN$ is used for dataset generation and then extracted starting at position $MN+{{L}_{CP}}+{{\theta }_{\text{d}}}+M{{\theta }_{\text{t}}}$, where $M$, $N$, and ${{L}_{CP}}$ are 256, 64, and 64, respectively. Therefore, the input signal containing STO can be represented as
\begin{equation}
\label{eq_5}
y[k] = r[k + {\theta _{\rm{d}}} + M{\theta _{\rm{t}}}].
\end{equation}

Since DL models require real-valued inputs, the complex-valued time-domain signal is split into real and imaginary components. The $i$-th input signal can then be represented as
\begin{equation}
\label{eq_6}
{y_i} = [{\mathop{\rm Re}\nolimits} \{ {y_i}\} ,{\mathop{\rm Im}\nolimits} \{ {y_i}\} ].
\end{equation}

The supervised learning classification label for frame synchronization corresponds to $\theta $. The number of classifications varies depending on the task scheme. This paper formulates OTFS frame synchronization as a time series classification problem, which can be divided into two schemes. One is to perform a one-stage classification sequence, where the label index range is consistent with the time series length. This scheme may cause a large computational load. The other is to first coarsely classify the time series of length $MN$ into $N$ categories based on STO, with each $M$ point corresponding to a label. After preliminary compensation of the sequence based on the results, iterative approximation estimation is performed, and the result is fed back into the model for re-estimation, with each $N$ point corresponding to a label. The latter label setting scheme is suitable for the coarse-to-fine ResNet frame synchronization model design of this paper and the superiority of this estimation method is analyzed in the performance evaluation.

Based on the above simulation data generation and pre-processing procedures, we constructed a multi-scenario OTFS dataset containing 90,000 samples (30,000 per channel type). To ensure rigor, 80\% of the samples (24,000 groups per channel) are randomly selected for training, while the remaining 20\% (6,000 groups) constitute the independent test set.

\subsection{Evaluation Results}
The performance of the proposed method is evaluated using accuracy and root mean square error (RMSE). Accuracy measures the proportion of samples where the estimated class is consistent with the true class, with a value range from 0 to 1. A value closer to 1 indicates more accurate frame synchronization. In this paper, accurate synchronization is recorded when the estimated frame start index is the same as the true frame start index. RMSE reflects the magnitude of the estimation error, where a value closer to zero indicates higher estimation accuracy.

In addition, this paper compares the performance of different conventional frame synchronization schemes for the single-pilot OTFS signal used in this experiment, such as the cross-correlation algorithm and the 2D autocorrelation algorithm based on pilot. To comprehensively evaluate the estimation performance of the ResNet model for coarse-to-fine synchronization, another model such as two-stage CNN architecture is also compared. Furthermore, to verify that the proposed method yields performance comparable to that of a one-stage classification method that performs for all $MN$ categories simultaneously.

\begin{figure*}[!t]
    \centering
    \subfloat[]{\includegraphics[width=0.8\columnwidth]{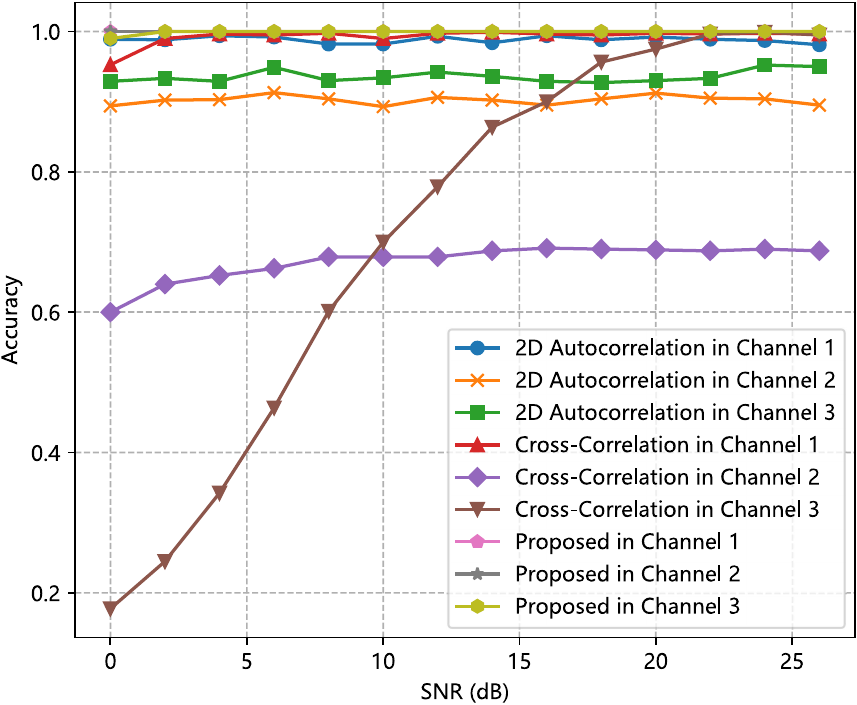}\label{fig_7a}} 
    \hfil
    \subfloat[]{\includegraphics[width=0.8\columnwidth]{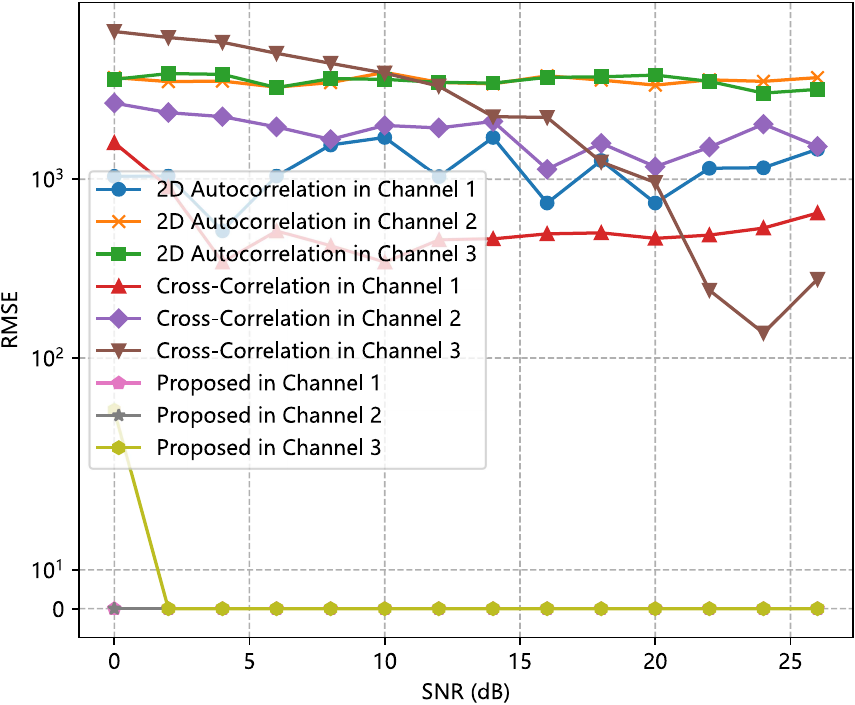}\label{fig_7b}}
    \caption{Performance of the proposed method and conventional synchronization schemes. (a) Accuracy. (b) RMSE.}
    \label{fig_7}
\end{figure*}

In \cref{fig_7}, we compare the performance of the proposed method and conventional synchronization schemes, such as cross-correlation algorithm and 2D autocorrelation algorithm. Among them, since the cross-correlation algorithm requires inserting a preamble into the signal, to ensure the fairness of the experiment as much as possible, the preamble length is set to 256 samples. The cross-correlation algorithm (Channel 2) struggles to achieve high synchronization accuracy, even at higher SNR. Despite achieving high accuracy, both the 2D autocorrelation (Channels 1-3) and cross-correlation (Channels 1 and 3) algorithms exhibit poor RMSE performance, resulting in significant STO. In contrast, the proposed synchronization method not only eliminates the need for inserting a known preamble (and its associated overhead), but also maintains high frame synchronization accuracy and low RMSE in Channels 1-3. This is because the 2D autocorrelation algorithm relies on the intrinsic periodicity of pilot rows to perform autocorrelation operations, while the cross-correlation algorithm depends on known preamble sequences for cross-correlation computations. Both approaches are susceptible to degradation under multipath interference and variations in SNR. In contrast, the proposed method leverages both the periodic characteristics of pilot rows and the overall structural features of the signal frame for inference, thereby exhibiting superior robustness.

\begin{figure*}[!t]
    \centering
    \subfloat[]{\includegraphics[width=0.8\columnwidth]{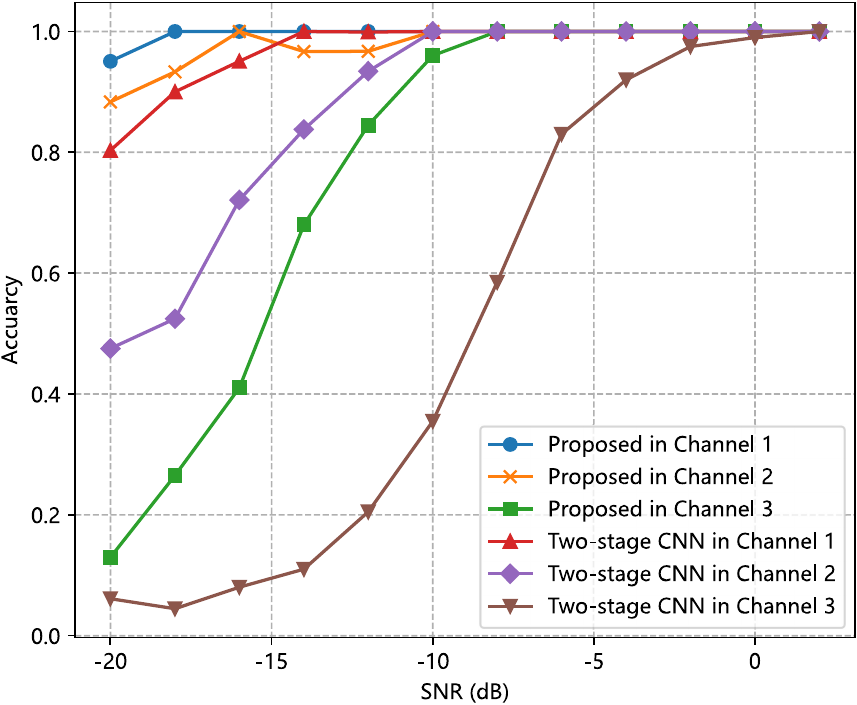}\label{fig_8a}} 
    \hfil
    \subfloat[]{\includegraphics[width=0.8\columnwidth]{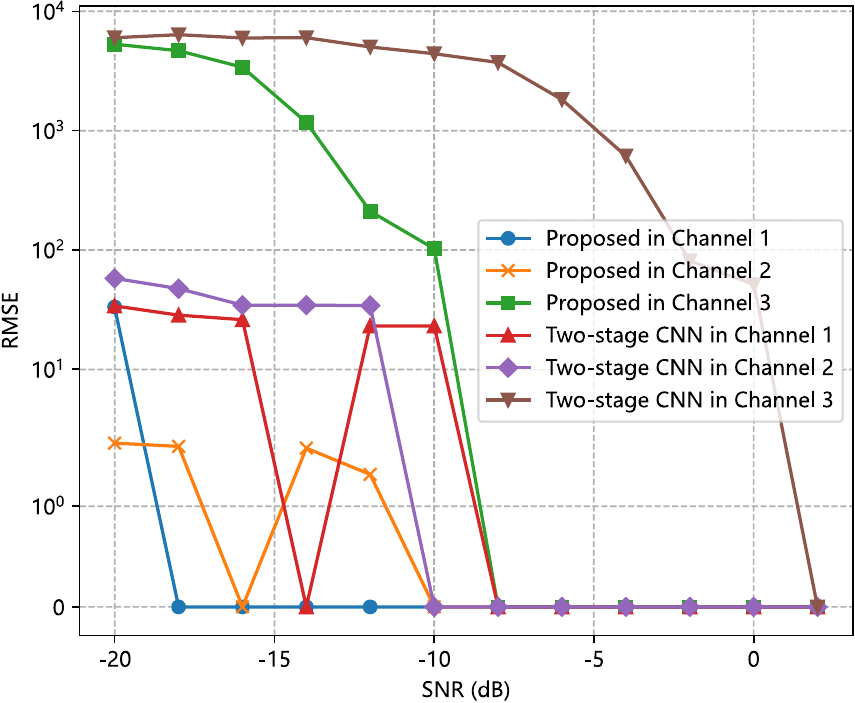}\label{fig_8b}}
    
    \caption{ Performance of the proposed model and CNN model. (a) Accuracy. (b) RMSE.}
    \label{fig_8}
\end{figure*}

To comprehensively evaluate the performance of the proposed ResNet model, a baseline model based on a standard CNN is introduced for comparison. As a well-established architecture extensively validated for time-series classification tasks, CNNs are frequently employed as a performance benchmark for evaluating novel models \cite{ref16}. To ensure a fair comparison, this CNN baseline model adopts an identical two-stage synchronization strategy to that of our proposed method and utilizes the same dataset, preprocessing steps, and training scheme. Due to the performance saturation at high SNRs, to focus ResNet's robustness under low SNR regimes and complex channels, we calculate the accuracy and RMSE of each model across a SNR range from $-20$ dB to $2$ dB. As shown in \cref{fig_8}~\subref{fig_8a}, synchronization performance is greatly affected by the SNR and the multipath fading, the lower the SNR, the lower the synchronization accuracy, and the simpler the channel, the higher the synchronization accuracy. Observing the RMSE is shown in \cref{fig_8}~\subref{fig_8b}. For Channel 1 and 2, it can be noted that under low SNR conditions, even if the frame start sampling moment is not perfectly estimated, the estimation range can still be narrowed as much as possible. Based on the impact of STO on the received signal and OTFS demodulation, this reduction in the STO estimation range also helps to reduce the degree of ICI and ISI. For Channel 3, the RMSE increases sharply as the SNR decreases, and the ResNet model shows an inflection point is about $-12$ dB, which is lower than that of the CNN model. This means that the ResNet model shows better robustness for the frame synchronization task under more complex channel conditions.

\begin{figure*}[!t]
    \centering
    \subfloat[]{\includegraphics[width=0.8\columnwidth]{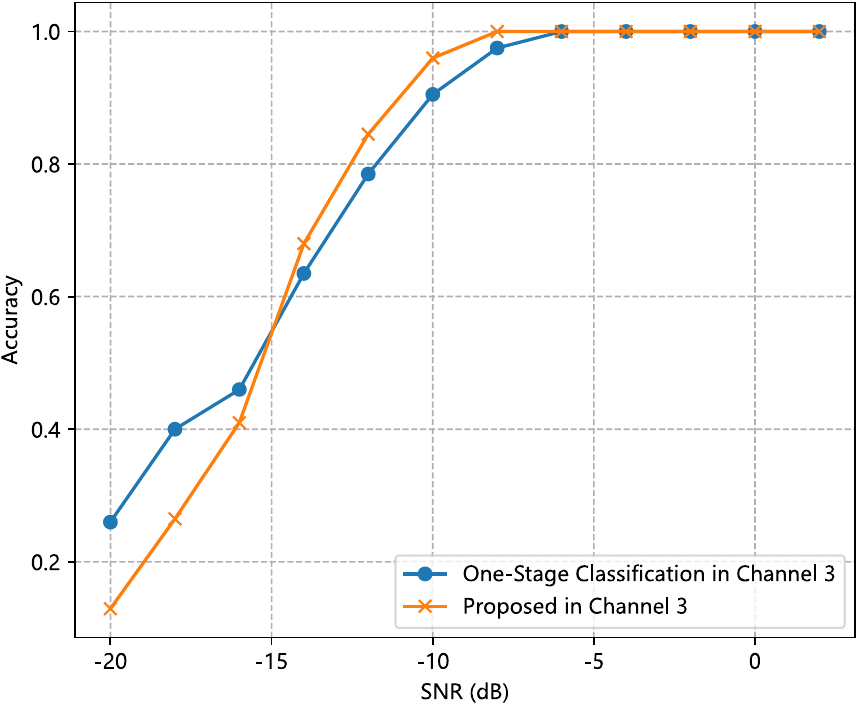}\label{fig_9a}} 
    \hfil
    \subfloat[]{\includegraphics[width=0.8\columnwidth]{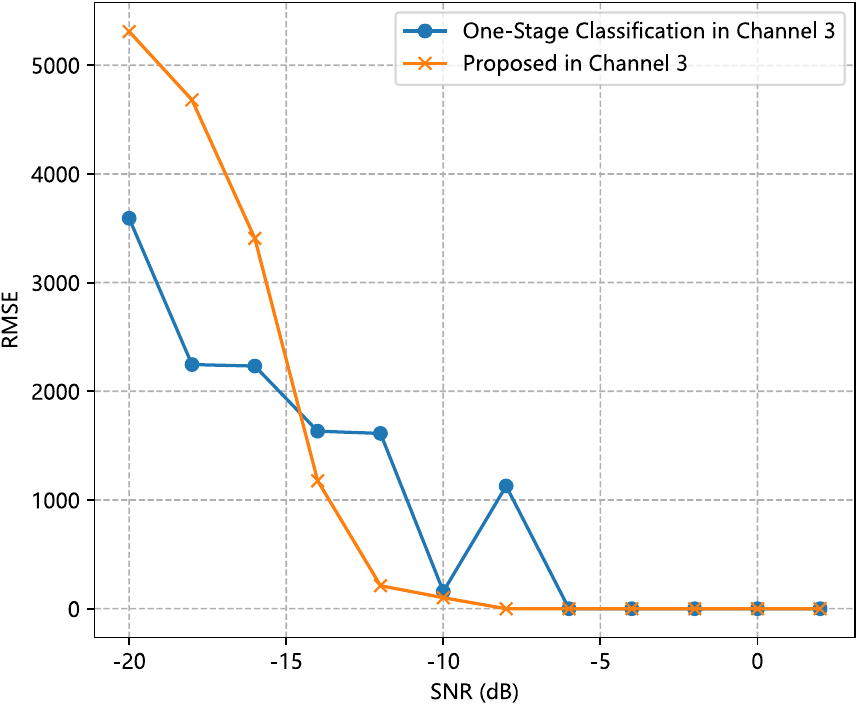}\label{fig_9b}}
    
    \caption{Performance of the proposed method and one-stage classification method. (a) Accuracy. (b) RMSE.}
    \label{fig_9}
\end{figure*}

\cref{fig_9} compares the RMSE and accuracy of the proposed method against a one-stage ResNet-based classifier designed for direct  classes classification. The results demonstrate that our two-stage, coarse-to-fine classification scheme achieves STO estimation performance comparable to the one-stage approach. Crucially, this comparable performance is attained with a marked reduction in both computational complexity and model dimensionality.

\subsection{Complexity Analysis}

Time complexity is quantified using floating point operations (FLOPs), which measure the computational cost of a single forward pass. In this paper, FLOPs are reported in millions (M) and calculated as twice the number of multiply-accumulate operations (MACs) to account for both addition and multiplication. This metric directly correlates with inference latency and resource demands; specifically, higher FLOPs imply increased processing time and energy consumption, presenting significant challenges for resource-constrained devices.

Moreover, space complexity is represented by the number of parameters, which indicates the total number of trainable parameters (weights and biases) in the model. It is a metric for measuring the size and storage requirements of a model, reflecting the model's capacity (ability to learn complex patterns) and memory usage. The space complexity in this paper is calculated in millions (M). The number of parameters determines the model's memory footprint, storage requirements, and training difficulty. More parameters increase the model's capacity but also increase the risk of overfitting and deployment costs.

For the conventional 2D autocorrelation and cross-correlation methods, as these are deterministic algorithms based on mathematical derivations. Thus, they require no training process. However, their time complexity is determined by the signal dimensions and the search window size. Specifically, the complexity of the 2D autocorrelation method scales with $\mathcal{O}(MN^2)$, while the cross-correlation method scales linearly with the product of the search window size of $MN$ and preamble length of $L_{\text{seq}}$, expressed as $\mathcal{O}(MN \cdot L_{\text{seq}})$.

We use floating point operations (FLOPs) to represent time complexity and the number of parameters to represent space complexity.
FLOPs are calculated in millions (M), representing the total number of operations per forward pass.
Space complexity is measured in millions of parameters (M).
For conventional methods (2D autocorrelation and cross-correlation), complexity is deterministic: 2D autocorrelation scales with $\mathcal{O}(MN^2)$, while cross-correlation scales with $\mathcal{O}(MN \cdot L_{\text{seq}})$.

\begin{table*}[!t]
\caption{Complexity Analysis of Different Methods\label{table_2}}
\centering
\begin{tabular}{|l||c||c||c|}
\hline
Method &  Time Complexity (FLOPs / M) &  Space Complexity (Parameters / M) &  Runtime / s\\
\hline
Proposed & 195.035 & 10.501 & 0.0232\\
\hline
Two-stage CNN & 188.108 & 10.494 & 0.0193\\
\hline
One-stage classification & 1205.726 & 537.431 & 0.0494\\
\hline
Cross-correlation & 33.554 & N/A & 0.0284\\
\hline
2D autocorrelation & 8.258 & N/A & 0.0194\\
\hline
\end{tabular}
\end{table*}

As shown in \Cref{table_2}, the time complexity and space complexity of different frame synchronization schemes are compared. Although conventional methods possess the lowest theoretical FLOPs, they lack the adaptability of learning-based methods. It can be seen from the table that the two-stage, coarse-to-fine synchronization based on ResNet proposed in this paper has significantly lower complexity compared to the one-stage classification model based on ResNet that classify the entire signal length as one dimension, without sacrificing the estimation performance of the model. Furthermore, the choice of using ResNet instead of CNN in this paper, combined with the analysis of synchronization performance in the previous experimental results, shows that using the ResNet model for coarse-to-fine synchronization does not significantly increase complexity compared to a two-stage CNN model. Therefore, the ResNet model can be considered the optimal model for single-pilot OTFS signal frame synchronization.

Furthermore, we record the runtime to execute inference on 1 input signal of each neural network method, as well as the conventional algorithms, in \Cref{table_2}, by evaluating them on an Intel Core i9-13905H CPU. Conventional algorithms such as cross-correlation and 2D autocorrelation demonstrate significant advantages in small-scale synchronization tasks executed on the CPU. It can also be seen that in our proposed method, even though utilizing the DL model makes the runtime increase, comparable real-time performance can still be maintained. Moreover, if GPUs are used to perform DL-based synchronization tasks, the reduction in runtime speed can be compensated and large-scale operations can be achieved.

\section{Conclusion}
In this paper, the frame synchronization problem for OTFS systems in high-mobility scenarios has been investigated, where the estimation of STO has been formulated as a DL-based classification task. To address the prohibitive computational complexity of direct one-stage classification, a novel non-data-aided coarse-to-fine synchronization method based on ResNet architecture has been proposed. By leveraging the periodic features of embedded pilots, this hierarchical architecture has effectively eliminated preamble overhead while significantly reducing the search space. Furthermore, we have constructed a dataset incorporating diverse channel conditions to facilitate robust model training and evaluation. Extensive simulation results have demonstrated that the proposed method has achieved superior synchronization accuracy and robustness compared to conventional algorithms and standard CNN baselines, particularly in challenging low SNR and high-mobility environments. Finally, complexity analysis has confirmed that the proposed scheme has realized a significantly more efficient inference process than one-stage classifiers without compromising estimation precision, thereby offering a promising low-complexity and high-performance solution for future wireless physical layer applications.

\bibliographystyle{IEEEtran}
\bibliography{mybib}

\vfill

\end{document}